\documentclass[aps,prl,twocolumn,amsmath,amssymb,superscriptaddress]{revtex4}
\usepackage{amssymb}
\usepackage{graphicx}
\usepackage{dcolumn}
\usepackage{verbatim}
\begin{document}


\title{First-principles modeling of multferroic RMn$_2$O$_5$}

\author{Kun Cao}
\affiliation{Key Laboratory of Quantum Information, University of
Science and Technology of China, Hefei, 230026, People's Republic of
China}

\author{Guang-Can Guo}
\affiliation{Key Laboratory of Quantum Information, University of
Science and Technology of China, Hefei, 230026, People's Republic of
China}

\author{David Vanderbilt}
\affiliation {Department of Physics and Astronomy, Rutgers University,
Piscataway, New Jersey 08854-8019, USA}

\author{Lixin He \footnote{Email address: helx@ustc.edu.cn}}
\affiliation{Key Laboratory of Quantum Information, University of
Science and Technology of China, Hefei, 230026, People's Republic of
China}

\date{\today}

\begin{abstract}

We investigate the phase diagrams of RMn$_2$O$_5$ via a first-principles
effective-Hamiltonian method. We are able to reproduce the most important
features of the complicated magnetic and ferroelectric phase transitions.
The calculated polarization as a function of temperature agrees very well with
experiments. The dielectric-constant step at the commensurate-to
-incommensurate magnetic phase transition is well reproduced.
The microscopic mechanisms for the phase transitions are discussed.

\end{abstract}
\maketitle

%
RMn$_2$O$_5$ (R=Tb, Dy, Ho, Y etc.) belong to a very special class
of multiferroics, because the ferroelectricity is driven by the
magnetic ordering\cite{erenstein06,cheong07,wang07}. These compounds
therefore possess strong magnetoelectric (ME) coupling, showing
remarkable new physical effects, such as the colossal
magnetodielectric \cite{hur04b} and magneto-polarization-flop
effects \cite{kimura03,goto04,hur04}, etc. The strong ME coupling
effects are not only interesting in the view of fundamental physics,
but also they have potential important applications in future
multifunctional devices.

Because of the complex magnetic interactions and the ME coupling,
RMn$_2$O$_5$ compounds undergo several magnetic and associated
electric phase
transitions\cite{chapon06,hur04b,higashiyama05,kimura09} upon
cooling from room temperature to near zero temperature. Generally,
these compounds transform at about 40\,K from a paramagnetic (PM)
phase to an antiferromagnetic (AFM) phase whose magnetic ordering is
initially commensurate (CM) along the $a$ axis. This phase
transition is accompanied by a ferroelectriclike transition, with
the appearance of a spontaneous polarizations and a divergence of
the dielectric constant. When the temperature is lowered further to
about 20\,K, the magnetic structures become incommensurate (ICM)
along the $a$ axis, and there is a drop of the electric polarization
together with the appearance of a step in the dielectric constant
\cite{hur04,higashiyama05,hur04b}. The special phase transition
sequence \cite{hur04,chapon04} is very puzzling and the driving
forces for the phase transitions are not understood. It is therefore
very important to explore the closely related magnetic and electric
phase transitions to gain a full understanding of the microscopic
mechanism of the ME coupling and novel physics in these materials.

Recent neutron scattering experiments
\cite{chapon04,radaelli09} as well as first-principles calculations
\cite{wang07,
wang08} suggest that the strong ME coupling in RMn$_2$O$_5$
is due to the ``exchange striction'' effect.
However, previous
first-principles calculations \cite{wang07,wang08}
were limited to zero temperature, and
did not provide information about the phase transitions.
The phase diagrams of RMn$_2$O$_5$ materials have been studied via a
phenomenological approach \cite{harris08}. This approach, based on
symmetry considerations only, does not reveal any of the microscopic
mechanisms of the ME coupling. In this letter, we
present a first study of the phase diagrams of RMn$_2$O$_5$ materials
as a function of temperature by using a first-principles
effective-Hamiltonian method \cite{zhong94}. We obtain the most important
features of the phase diagram, including the the magnetic PM-CM-ICM
transitions, the accompanying ferroelectric transitions, the electric
polarization as function of the temperature, and the dielectric-constant
step at the CM-ICM transition.

\begin{figure}
\centering
\includegraphics[width=2.0in]{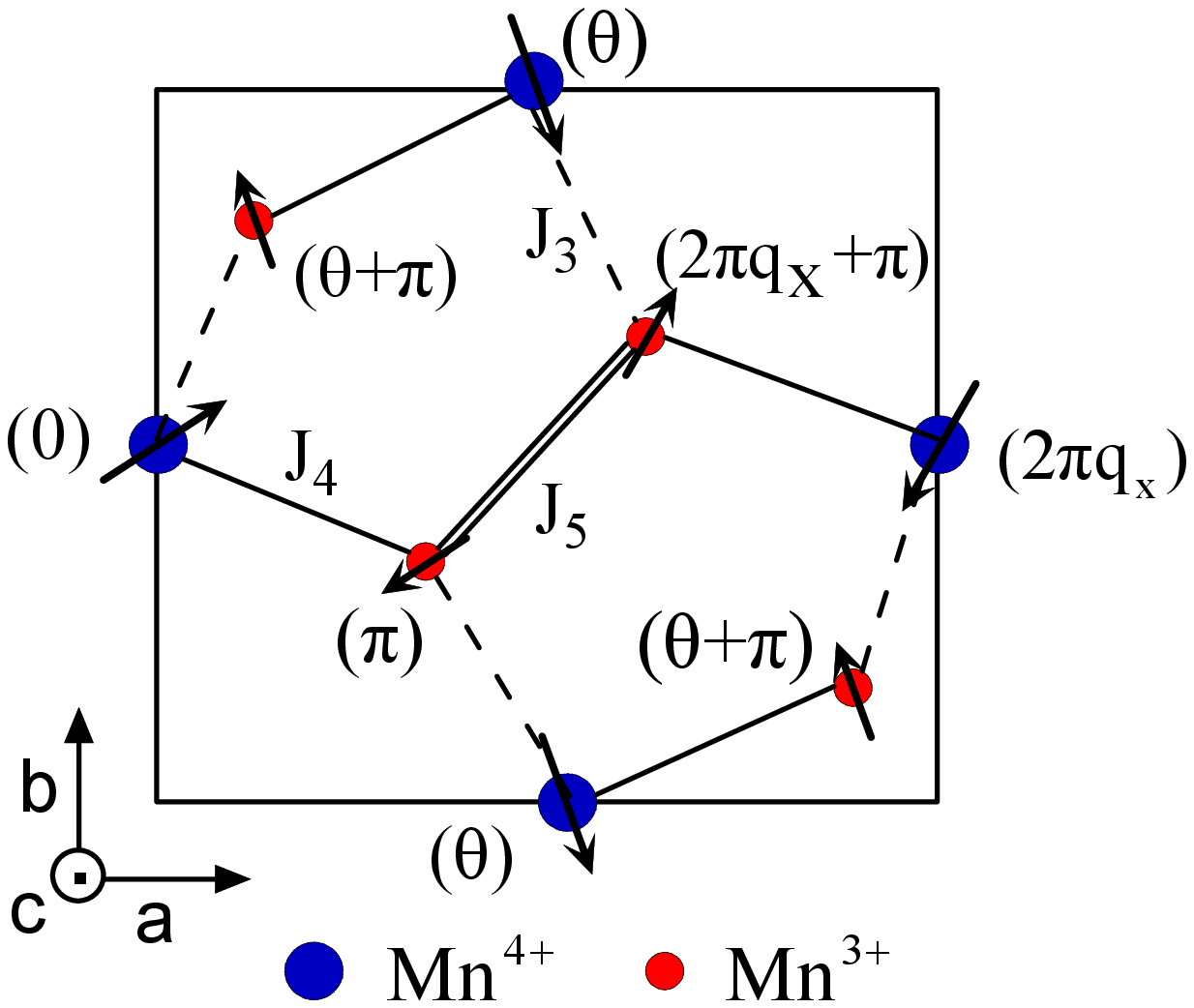}
\caption {(Color online) A schematic sketch of the magnetic structure
projected onto the $ab$ plane. The angle of spin
orientation is given in the parentheses.  The dashed,
solid, and double lines represent $J_3$, $J_4$ and $J_5$ exchange interactions
respectively. $J_1$ and $J_2$ (not shown) are along the $c$-direction.}
\label{fig:ICM}
\end{figure}

%
The high-temperature crystal structure of TbMn$_2$O$_5$ is
orthorhombic (space group Pbam)
with four TbMn$_2$O$_5$ formula units
per primitive cell, containing  Mn$^{4+}$O$_6$ octahedra
and Mn$^{3+}$O$_5$ pyramids \cite{alonso97}.
The effective Hamiltonian was derived in Ref.~\onlinecite{wang08} from a
Heisenberg-like model.
The
spin-phonon coupling comes from the dependence of the exchange
interactions $J_{\alpha}$ on the phonon modes ${u}_{\lambda}$.
$J_{\alpha}(\{{u}_{\lambda}\})$ was expanded around
the high-symmetry structure 
to second order in the phonon mode amplitudes.
Five nearest-neighbor (NN) exchange interactions were included, as
sketched in Fig.~\ref{fig:ICM}.
$J_3$ is the Mn$^{4+}$-Mn$^{3+}$ superexchange interaction
through pyramidal base corners, while $J_4$ is the superexchange
interaction through the pyramidal apex \cite{chapon04}.
The Mn$^{3+}$ ions in connected pyramids couple to each other
antiferromagnetically through $J_5$, whereas
$J_1$ and $J_2$ couple Mn$^{4+}$ ions along the $c$ axis.

At lower temperature a further distortion occurs, reducing the
crystal symmetry to Pb2$_1$m.
The lattice distortion involves 14 IR-active $B_{2u}$ modes \cite{wang07,wang08}.
Since the symmetry-lowering displacement
is extremely small, we treated this displacement
(henceforth $u$) as the only phonon normal mode in the model.
Only the single parameter
$J_3'$=${\partial J_3 / \partial u}$
was assumed to be involved in the first-order spin-phonon interaction.
The neglect of $J_{\alpha}''$ terms, which renormalize the phonon
frequencies and lead to the phonon anomalies near the magnetic phase
transitions \cite{flores06,cao08,shen08}, is justified because these
terms have a quite small effect on the phase diagrams studied here.
The simplified Hamiltonian is then
\begin{eqnarray}
\label{eq:eff_h3}
E(\{u_{k} \}) &=& E_0+\sum_{k} {1\over 2} m \omega^2
u_{k}^{2} + \sum_{k\ne l} {1\over 2} \xi_{kl} u_{k}u_{l}
 \\ \nonumber
&& -\sum_{ij \in {J_{\alpha}} }J_{\alpha}(0) {\bf
S}_i\cdot {\bf S}_j -\sum _{ij\in {J_3}}\sum_{k}
J_{3}' \, u_{k} {\bf S}_i\cdot {\bf S}_j\, .
\end{eqnarray}
Here $E_0$ is the energy of the high symmetry structure without
magnetic interactions, whereas $m$ and $\omega$ are the reduced mass
and frequency of the IR-active mode,
${u}_k$ is the $k$-th local phonon mode , and  $\xi_{kl}$ are
force-constant matrix elements that couple the NN local phonon
modes.
This last term was absent from Ref.~\onlinecite{wang08}, but is
included here to describe the phonon dispersion properly.
We assume that the $\xi_{kl}$ are isotropic in the
$ab$ plane, and we neglect the much smaller couplings along the
{\it c} direction.

%
Since the RMn$_2$O$_5$ compunds have similar phase diagrams, we
choose TbMn$_2$O$_5$ as an example, and determine the
parameters for the simplified Hamiltonian Eq.~(\ref{eq:eff_h3})
by carrying out a series of first-principles calculations on
this compound \cite{wang07,wang08}.
The calculations were based on density-functional
theory within the generalized-gradient approximation
(GGA) implemented in the Vienna Ab-initio Simulations Package
(VASP)\cite{kresse93,kresse96}. Projector augmented-wave (PAW)
pseudopotentials\cite{blochl94} and a 500\,eV plane-wave cutoff were
used.
Spin polarization was included in the collinear approximation.
The resulting $J$ parameters can nevertheless be used
to model noncollinear situations.

To get the spin-phonon coupling constant $J_{3}^\prime$,
it is enough to use the the energy difference $\Delta E$ between the
high-symmetry and the ground-state low-symmetry structures.
To simplify the notation, we redefine $u$ to be a dimensionless
parameter taking the value of unity at the ferroelectric low-symmetry
state, and assign spin moments $|{\bf S}_i|$=1.0 as well.  Then
it is easy to show that $J_3^\prime=\Delta E/4$.
We have $J_3^\prime \sim$ 1.125\,meV.
In order to
calculate the phonon coupling constant $\xi_{kl}$, we calculate the total
energies of different local-mode configurations.
In practice, we find that including the short-range
phonon interaction only has a very small effect on the results.
The exchange interactions $J_1$ -  $J_5$ were fitted
to the total energies of different spin configurations
and were given in Ref.~\onlinecite{wang08}.
Alternatively, the exchange interactions can be calculated from the
extended Kugel-Khomskii model \cite{das08}.

In the present work, we have now also
fitted the parameters to GGA+U calculations \cite{giovannetti08}
with 1.0\,eV\,$\le$\,$U$\,$\le$\,4.0\,eV on the Mn ions.
We find that $J_3^\prime$ decreases with increasing U,
falling to $J_3^\prime\,=\,0.325$meV at $U\,=\,4.0$\,eV.  As we shall
see, this improves the comparison of some of our later simulation
results with experiment.  Unfortunately, increasing $U$ also
worsens the agreement with experiment for the
$J$ parameters themselves.  This tension between the fitting of
$J$ and $J'$ parameters will be further discussed later.

%
We investigated the finite-temperature behavior of our
effective Hamiltonian by using Monte Carlo (MC)
simulations. Traditional serial-temperature MC methods have great
difficulty treating
systems with complex frustrated interactions.
Moreover, the present system has a first-order CM-AFM--to--ICM
phase transition which would be very difficult to treat
using conventional methods. Here we adopt the replica-exchange method
\cite{swendsen86} in which one simulates $M$ replicas each at a different
temperature $T$ covering a range of interest, and allows configurational
exchange between the replicas.  Importantly, the inclusion of high-$T$
configurations ensures that the lower-$T$ systems can access a broad
phase space and avoid becoming trapped in local minima.

We perform the simulations on an $L \times L \times
L$ cubic cell with periodic boundary conditions. Each unit cell
contains eight spins and two local phonon modes. In the simulations,
one MC sweep is defined to consist of a series of attempts of all
variables.  We performed the simulations at
temperatures ranging from 3 to 90\,K. The
temperatures are adjusted to ensure that the exchange rates between adjacent
remain close to 20\%.
%
At each $T$ we carry out an initial 10$^4$ sweeps to prepare the
system before allowing replica exchange.  We discard these, as well
as the first $10^6$ sweeps after replica exchange is started, when
computing equilibrium properties. Sample averages are accumulated
over 2$\times 10^6$ sweeps, without replica exchange to avoid a sign
problem.

\begin{figure}
\centering
\includegraphics[width=2.5in]{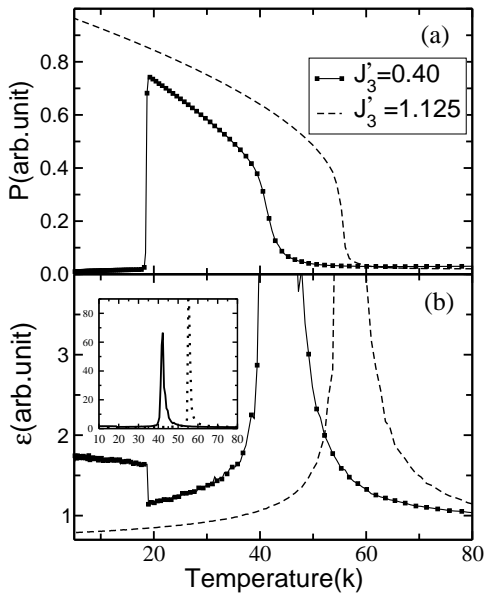}
\caption {(a) The electric polarization P as a function of
the temperature;
(b) The dielectric constant $\varepsilon$ as a function of temperature
for both $J_3^\prime=1.125$ and $J_3^\prime=0.4$.
The inset window shows the full view of $\varepsilon$. The dielectric
constants are normalized to unity at high temperature.
}
\label{fig:polar}
\end{figure}

%
We give here the results of typical simulations on a
12$\times$12$\times$12 cell.
Figures~\ref{fig:polar}(a) and (b) depict the polarization and dielectric
  constant respectively,  which are calculated via $P$=$\langle u \rangle$ and
$\epsilon$=$(\langle u^2 \rangle -  \langle u \rangle^2)/T$.
If we use $J_3^\prime=1.125\,$meV and the exchange interactions
are fitted from the GGA calculations, we get a
single magnetic PM--to--CM-AFM transition at about 58\,K,
accompanied by a ferroelectric transition
(shown as the dotted lines in Fig.~\ref{fig:polar}). This result misses the
important CM-to-ICM phase transition and overestimates the PM-to-CM transition
temperature.  The problem can be traced to the too-large spin-lattice
coupling constant $J_3^\prime$.
Including the on-site Coulomb $U$ can reduce $J_3^\prime$, but at
the same time it worsens the exchange interactions.
We thus find that neither an effective Hamiltonian built on a pure GGA
calculation, nor one built on GGA+U with a single value of $U$,
can give good overall agreement with experiment.

However, if we are willing to adjust the parameters by using
the exchange coupling taken from pure GGA and the $J_3^\prime$
from GGA+U, the situation improves dramatically.
If $J_3^\prime$ is reduced to about 0.4\,meV, as obtained from GGA+U
with $U$=3\,eV, we obtain two phase transitions at about 42 and 18\,K
respectively, in very good agreement with experiment. The
nature of each phase transition was identified via Fourier analysis of
the spin configurations.  At 80\,K, the spins are fully disordered,
indicating a PM phase. When $T$ is lowered to 25\,K, the spin spectrum
shows a dominant peak at ${\bf q}=(0.5,0,0.5)$, suggesting
a CM-AFM phase. The Fourier spectrum of the spin structure at 5\,K
shows dominant peaks at ${\bf q}$=(5/12,0,0.5), indicating it is in the
ICM phase.  We therefore obtain the most important PM-CM and CM-ICM
phase transitions, and the transition temperatures agree very well with
the experimental values
($\sim$38-44\,K for the AFM-CM ordering along
the $a$ axis, and $\sim$20\,K for the ICM ordering)
\cite{chapon06,hur04b,higashiyama05,kimura09}.
The calculated $q_x$=5/12 is slightly smaller than the experimental
values ($\sim$0.46-0.48).
It is worth noting that the calculated $q_x$ is restricted by the
supercell sizes in the simulation, which can be improved by increasing
the supercell size. Simulations on a 14$\times$14$\times$14 cell
give $q_x$=3/7.  While we did not reproduce the correct $q_z$ in both
the CM and ICM phases (probably because we only have NN
interactions in the model Hamiltonian),
the fact that we nevertheless reproduce the correct phase transition sequence
tends to confirm that the $q_z$ value is not important for the ME coupling in
these materials \cite{chapon04,wang07}.

The solid curves in Figs.~\ref{fig:polar}(a) and (b) show the
spontaneous polarization $P$ and the dielectric constant $\epsilon$
as functions of $T$ for $J_3'$=0.4\,meV.  The polarization increases
strongly as the temperature is reduced through the
PM-CM transition, but then it drops suddenly almost to zero
at the CM-ICM transition.
The magnetically induced polarization behaves as
$P \propto \langle {\bf S}_3 \cdot {\bf S}_4 \rangle$, where
${\bf S}_3$ and ${\bf S}_4$ are the spins of the Mn$^{3+}$ and
Mn$^{4+}$ ions coupled via the $J_3$ interaction.
In the ICM phase,  ${\bf S}_3$ and ${\bf S}_4$ are almost orthogonal
(i.e., $\theta \sim \pi/2$ in Fig.~\ref{fig:ICM}), whereas in the CM phase
they are parallel or antiparallel.
These results are in excellent agreement with the experimental
results for RMn$_2$O$_5$ compounds
\cite{chapon04,higashiyama05,kagomiya03,hur04b},
especially for YMn$_2$O$_5$ \cite{kagomiya03} and HoMn$_2$O$_5$
\cite{higashiyama05}.
Note, however, that our simulation does not reproduce the
reemergence of a polarized state observed experimentally in
TbMn$_2$O$_5$ at still lower temperature \cite{hur04}.
This is probably because we ignore the spins of Tb 4$f$ electrons
in our model. Experimentally, it is observed that Tb is
magnetically ordered below $\sim$10\,K, which might play an important
role in the reemergence of the polarization at low $T$
\cite{hur04}.

The dielectric constant shows a peak at the PM-CM transition as a
consequence of the ferroelectric phase transition. Most
interestingly, the dielectric constant step at the CM-ICM transition
has been well reproduced in the simulation, in which $\varepsilon$
jumps by about 75\% in going from the CM phase at 18\,K to the ICM
phase.
The step is very interesting and important, because it may directly
relate to the colossal magnetodielectric effect, which happens
just at the CM-ICM transition temperatures in these materials.

\begin{figure}
\centering
\includegraphics[width=2.5in]{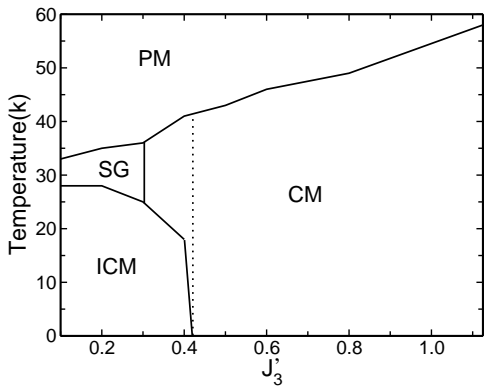}
\caption {The $J_3^\prime$-temperature phase diagram of the system, where PM,
  CM, ICM and SG represent paramagnetic, commensurate, incommensurate
and spin-glass-like phase respectively.} \label{fig:phase}
\end{figure}

To gain a better understanding of the spin-lattice coupling effects on
the magnetic and structural phase transitions, we plot the
$J_3^\prime$-temperature phase diagram in Fig.~\ref{fig:phase}. As we
see, if the spin-phonon coupling is too strong ($J_3' >$ 0.42\,meV),
there is only a PM-CM transition, and no CM-ICM transition
(as in BiMn$_2$O$_5$ \cite{munoz02}).
In contrast, if $J_3^\prime$ is very
small ($J_3' <$ 0.30\,meV), the CM state will not appear, and instead a
state having spin-glass (SG) character will appear above the ICM state. The
complex nature of the phase diagram is due to the frustration of the
$J_3$ interactions. It is easy to see that the CM states do not have
the lowest magnetic energies, since $J_3$ induces spins to
rotate to decrease the energy.
Since the spins have the same wave vector $q_z=0.5$ along the $c$ axis
in both the CM and ICM phases,
the interactions due to $J_1$ and $J_2$ do not change in the
the two phases, and can be neglected in the discussion.
According to the phase factors shown in
Fig.~\ref{fig:ICM}, the energy of the ICM state
can be written as
\begin{equation}
E_{ICM} \approx 8J_4-2J_5\cos(2\pi q_x) +4J_3 [\cos\theta +
\cos(2\pi q_x-\theta)] \, .
\end{equation}
Here we assume that two spins connected by
$J_4$ are always antiparallel to each other, because the $J_4$
interactions, each having two NNs, are much stronger than the $J_3$
and $J_5$ interactions. We also ignore the phonon contribution,
because in the ICM phase, $\theta \sim \pi/2$ and the
energy from spin-lattice coupling $J_3'$ is small.
For the CM state, we have
$\theta=0$, $q_x=0.5$. Therefore,
\begin{equation}
E_{CM}= 8J_4+2J_5- 4J_3^\prime \, .
\end{equation}
The energy difference between the CM and ICM phases is determined by
the competitions among $J_3$, $J_5$ and $J_3^\prime$. In the case of
$J_3^\prime$=1.125\,meV, the energy of the CM state is always lower
than that of the ICM state,  and there is no CM-ICM transition.
However, when $J_3^\prime$ decreases to 0.4\,meV, one can find
suitable $\theta$ and $q_x$ that allow the ICM state be the ground
state. Since there is no group-subgroup symmetry relation between
the CM and ICM states, the phase transition between them is
necessarily a first-order one. We speculate that the transition
occurs because the entropy of the CM state is larger than that of
the ICM state for suitable $J_3^\prime $.

The above results are calculated from the $L$=12 cell. We have also
obtained similar results for the $L$=10 and 14 cells. However, due to the
subtle nature of the ICM state, the parameters $J_3$ and $J_3^\prime$
have to be slightly adjusted to produce results that are in good
agreement with experiments for different cell sizes.

The ``semi-empirical'' philosophy we have adopted here has been to
start with first-principles derived parameters, make the
minimal empirical modifications to the parameters to get good
agreement with experiment, and then use the resulting model to
make predictions.  The fact that we have to adjust $J_3^\prime$
by hand (through the choice of $U$) to obtain good agreement with
experiment is clearly somewhat unsatisfactory.  However, there
is considerable precedent for such an approach.  For example, in
simulations of ferroelectrics via effective-Hamiltonian methods,
it is a common practice to adjust the lattice constant to agree
with experiment through the application of a fictitious negative
pressure \cite{zhong94}.

In summary, we have investigated the phase diagrams of RMn$_2$O$_5$
using a first-principles effective Hamiltonian method.
We obtained the most important features of the phase diagrams of
multiferroic RMn$_2$O$_5$ compounds,
including the sequence of magnetic and ferroelectric
phase transitions.
Most importantly, we obtained the dielectric-constant step at
the commensurate-to-incommensurate magnetic phase transition, which is
key to understand the colossal magnetodielectric effects.
The work further clarified the microscopic mechanism of the magnetoelectric coupling
in RMn$_2$O$_5$, and can be useful for exploring other multiferroic materials.

L.H. acknowledges the support from ``Hundreds of Talents''
program from CAS, and NNSF of China, Grant No. 10674124. D.V.
acknowledges the support from NSF Grant DMR-0549198.


\end{document}